\documentclass[fleqn,usenatbib]{mnras}

\usepackage{newtxtext,newtxmath}

\usepackage[T1]{fontenc}

\DeclareRobustCommand{\VAN}[3]{#2}
\let\VANthebibliography\thebibliography
\def\thebibliography{\DeclareRobustCommand{\VAN}[3]{##3}\VANthebibliography}

\usepackage{multirow}
\usepackage{booktabs}
\usepackage{graphicx}	
\usepackage{amsmath}	

\title[Continuous gravitational waves from slowly spinning neutron stars]{Searching for continuous gravitational waves from slowly spinning neutron stars with DECIGO, Big Bang Observer, Einstein Telescope and Cosmic Explorer}

\author[G. Pagliaro et al.]{
Gianluca Pagliaro,$^{1,2}$\thanks{E-mail: gianluca.pagliaro@aei.mpg.de}
Maria Alessandra Papa,$^{1,2}$\thanks{E-mail: maria.alessandra.papa@aei.mpg.de}
Jing Ming,$^{1,2}$
and Martina Muratore$^{3}$
\\
$^{1}$Max Planck Institute for Gravitational Physics (Albert Einstein Institute), Callinstr. 38, 30167 Hannover, Germany\\
$^{2}$Leibniz Universit\"at Hannover, D-30167 Hannover, Germany\\
$^{3}$Max Planck Institute for Gravitational Physics (Albert Einstein Institute), D-14476 Potsdam, Germany
}

\date{Accepted XXX. Received YYY; in original form ZZZ}

\pubyear{\the\year{}}

\begin{document}
\label{firstpage}
\pagerange{\pageref{firstpage}--\pageref{lastpage}}
\maketitle

\begin{abstract}
We consider stably rotating highly magnetised neutron stars and glitching pulsars. We discuss the prospects for detecting continuous gravitational waves from these sources below 20 Hz with next-generation ground-based facilities such as the Einstein Telescope and Cosmic Explorer and space-based observatories such as DECIGO and Big Bang Observer. We demonstrate that these constitute interesting science targets. We use a robust sensitivity estimation method for future searches based on demonstrated performance. We show that the spin-down upper limit on the gravitational wave amplitude of all highly magnetised pulsars and magnetars suitable for a years-long fully coherent search, exceeds the smallest gravitational wave amplitude estimated detectable with DECIGO and Big Bang Observer. We find that the hidden magnetar candidate PSR J1852+0040 can be detected by Cosmic Explorer if it is emitting at least at 20\% of its spin-down luminosity. Finally, post-glitch transient continuous gravitational waves from magnetars are an interesting target for deci-Hz detectors, with all but one of the recorded glitches giving rise to a spin-down limit signal {\it{above}} the smallest detectable level.
\end{abstract}




\section{Introduction}
\label{sec:intro}
Non-axisymmetrically deformed spinning neutron stars source continuous gravitational waves. The frequency $f$ of a continuous wave produced by a sustained mass-quadrupole in a rotating neutron star occurs at two harmonics of the rotation frequency $\nu$: $f \! = \! \nu$ and $f \! = \! 2 \nu$ (see \citet{2023APh...15302880W} or \citet{2023LRR....26....3R} for recent reviews on the topic). From this we immediately see that ground-based detectors are not optimally suited to detect emission from neutron stars spinning below 10 Hz.

There are many ways in which a neutron star can be deformed; for instance due to crustal temperature anisotropies \citep{1998ApJ...501L..89B, 2000MNRAS.319..902U}, magnetic confinement of accreted matter \citep{1983AA...128..369H, 1998ApJ...496..915B}, starquakes \citep{2022MNRAS.511.3365G}, and strong internal magnetic fields
\citep{1953ApJ...118..116C, 1954ApJ...119..407F, 1984PhLA..103..193G, 1989MNRAS.239..751K}.

The presence of magnetic fields alters the hydromagnetic equilibrium configurations of neutron stars, distorting their geometry. 
Magnetic deformations mainly depend on strength and topology of the field inside the star and scale linearly with the magnetic field energy \citep{1953ApJ...118..116C}, making highly magnetised neutron stars potentially the most deformed.

Let's consider a non-axisymmetric neutron star: 
\begin{equation}
\label{eq:ellipticity}
\varepsilon=\frac{I_{xx}-I_{yy}}{I_{zz}} \neq 0 ,
\end{equation}
with $I$ being the moment of inertia tensor and $\hat{z}$ being along the star angular moment. The quantity $\varepsilon$ is known as the equatorial ellipticity (or simply ellipticity) of the neutron star. 
Magnetically induced deformations are axisymmetric with respect to the axis of symmetry of the magnetic field. 
If the latter is inclined at an angle $\alpha$ with respect to the spin axis (misaligned rotator configuration), there is a net non-axisymmetry that scales as follows
\begin{equation}
\label{eq:ellipticityVsB}
\varepsilon \propto \textrm{B}^2 \sin^2 \alpha,
\end{equation}
with B being the neutron star's internal magnetic field \citep{1996AA...312..675B}.
In the case of neutron stars with superconducting cores, the quantity $\textrm{B}^2$ on the right hand side has to be replaced by $\textrm{B} \cdot \textrm{H}_c$, where $\textrm{H}_c$, typically of order $\approx 10^{15}\,\textrm{G}$, is the highest magnetic field value below which superconductivity is maintained in the core \citep{2008MNRAS.383.1551A}.

The intrinsic amplitude of a continuous gravitational wave signal at a distance $D$ from the source is 
\begin{equation}
\label{eq:h0}
h_0=\frac{4\pi^2 G}{c^4} \frac{I_{zz}\varepsilon f^2}{D}.
\end{equation}
This shows that, at fixed $\varepsilon$, low-frequency signals are weaker, hence harder to detect, but together with Eq.~\ref{eq:ellipticityVsB} it also suggests that, for magnetic deformations, this may be compensated if the magnetic field is strong enough.

Highly magnetised neutron stars are not expected to be fast rotators: under magnetic dipolar braking, the kinetic energy loss is faster for stronger magnetic fields.
Since this is the primary mechanism through which rotating neutron stars spin-down, highly magnetised neutron stars are unlikely to be found spinning above 10 Hz. 
This is also the reason why a detection by current ground-based detectors of continuous waves from a magnetically deformed neutron star is virtually impossible \citep{2023ApJ...952..123P}. 
Observational evidence of the most strongly magnetised neutron stars, i.e. magnetars, also confirms that these objects are all slow rotators. We make this our working assumption even though scenarios have been proposed which could imply high magnetic fields and fast rotations \citep{2016MNRAS.459.3407S, 2023ApJ...952L..21S}.

In this paper we study the prospects for detection of continuous gravitational waves emitted by highly magnetised neutron stars with future ground-based and space-based observatories in the deci-Hz range. 

We also consider the case of post-glitch continuous gravitational wave emission from pulsars. We take an approach similar to that taken by \cite{2023MNRAS.519.5161M} but apply it to the deci-Hz frequency range. This is interesting because most  known glitching pulsars spin at frequencies in the deci-Hz region, making  them valid science goals for detectors such as DECIGO and Big Bang Observer.

The paper is organised as follows: Section \ref{sec:detectors} is dedicated to the description of the detectors considered and their sensitivity. In Sections \ref{sec:known_pulsars} and \ref{sec:cco_snr} we treat the case of stable continuous wave sources and their detectability, respectively for the case of known pulsars and magnetars, and central compact objects in supernova remnants.
In Section \ref{sec:glitch_source} we discuss the case of glitching pulsars. We draw our conclusions in Section \ref{sec:discussion}.

\section{Detectors and sensitivity}
\label{sec:detectors}
The next two decades will see the blooming of gravitational wave astronomy, with facilities observing from sub mHz to the kHz region: the 3rd generation Einstein Telescope \citep{2010CQGra..27s4002P} and Cosmic Explorer \citep{2019BAAS...51g..35R} will scan the few Hz - kHz range from the ground; space-based detectors -- immune to the seismic noise plaguing the ground-based facilities \citep{KD_1996} -- will cover the mHz region: LISA \citep{2024arXiv240207571C}, TianQin \citep{2016CQGra..33c5010L} and Taiji \citep{2020IJMPA..3550075R}; the O(0.1) Hz - O(1) Hz range will be covered by lunar-based seismometers \citep{2021ApJ...910....1H, 2023SCPMA..6609513L}, and space-based laser interferometers: DECIGO \citep{2006CQGra..23S.125K} and Big Bang Observer (BBO, \citet{2005PhRvD..72h3005C}).

Relevant to our study case are Einstein Telescope and Cosmic Explorer, and the space-based future observatories DECIGO and Big Bang Observer.
We do not consider lunar-based seismometers because the sensitivity of space-based laser interferometers operational in the same frequency band is more competitive, and we do not consider milli-Hz detectors because of the too low frequency coverage, more suitable for example to spinning white dwarfs \citep{2019MNRAS.490.2692K, 2020MNRAS.492.5949S, 2024MNRAS.531.1496S}.

For the Einstein Telescope we use the sensitivity estimations of \citet{2011CQGra..28i4013H} labelled as ``ET-D" in equilateral triangle configuration. At the time of writing the design of ET has not been finalised and configurations other than the ``ET-D" are also on the table, most notably two distant L-shaped detectors \citep{2023JCAP...07..068B}. For Cosmic Explorer we use the single detector 40 km “baseline” configuration defined in \citet{2022ApJ...931...22S}.
For DECIGO and Big Bang Observer we take the sensitivity estimates from Equations (5) and (6) of \citet{2011PhRvD..83d4011Y}, which are in turn obtained by considering the design parameters of \citet{2006CQGra..23S.125K} for DECIGO and of \citet{2006PhRvD..73d2001C} for Big Bang Observer.

\subsection{Detectability of continuous gravitational waves}
\label{sec:sensitivity}
Searches for continuous gravitational waves are characterised by their sensitivity depth \citep{2015PhRvD..91f4007B}, defined as
\begin{equation}
\label{eq:sens_depth}
\mathcal{D}^{\mathcal{C}} \coloneq \frac{\sqrt{S_h(f)}}{h_0^{\mathcal{C}}} \, [\textrm{Hz}]^{-1/2}
\end{equation}
where $\sqrt{S_h}$ is the amplitude spectral density, describing the sensitivity of a given detector, and $h_0^{\mathcal{C}}$ is the amplitude of the weakest signal we are able to detect at the confidence level $\mathcal{C}$ (tipically $\mathcal{C} = 90 \text{-} 95 \%$).
Despite the frequency dependence of $S_h$, it turns out that the sensitivity depth of a search is nearly constant at all frequencies. 

Let's now suppose that a search X has been carried, say using a certain amount of data, and that the sensitivity depth of that search is $\mathcal{D}_\textrm{X}$. If one planned to repeat that search, using the same amount of data but with a detector of the same type and a different sensitivity $S_h^{\textrm{new}}$, one could predict the weakest detectable signal from the new search, simply by solving Eq.~\ref{eq:sens_depth} with $S_h^{\textrm{new}}$ and $\mathcal{D}_\textrm{X}$:
\begin{equation}
\label{eq:h0VsSensDepth}
h_0^{\textrm{new}}(f)= \frac{\sqrt{S_h^{\textrm{new}}(f)}}{\mathcal{D}_\textrm{X}}.
\end{equation}

In this paper we will imagine repeating a search that has already been carried out but using data from future detectors, with their projected amplitude spectral densities. Using Eq.~\ref{eq:h0VsSensDepth} we can estimate the weakest detectable signal by that search. We will then consider a signal detectable when its amplitude $h_0$ is at least as large as the smallest detectable signal. This approach leads to a robust sensitivity estimate and it has been widely used in recent works, for example by \citet{2023ApJ...952..123P}, \citet{2023MNRAS.519.5161M} and \citet{2023arXiv231012097T}. 
We want to use this approach to estimate the detectability of continuous signals by both 3rd generation ground-based detectors and space-based deci-Hz detectors, under the conservative assumption that we simply repeat a search that has been performed using current ground-based detector data. 

The sensitivity depth folds-in the response to continuous waves of a specific type of detector, and so adopting the sensitivity depth from LIGO-like searches for space-based detector data might not be correct. Also, the difference in the total  number of detectors must be taken into account. 

\begin{itemize}
\item The signal as seen through a gravitational wave detector is amplitude-modulated by the antenna beam-pattern function moving across the sky as the detector moves. Ground-based detectors have a different antenna beam-pattern function than space-based ones, and also move differently: 
while ground-based detectors move like a point on the surface of the Earth, both DECIGO and Big Bang Observer will move in a heliocentric Earth-trailing orbit, just as LISA (see \citet{earth_trail_motion} for LISA's motion visualisation). This difference impacts the optimal signal-to-noise ratio (Section 3.3 of \citet{Jaranowski_1998}) that can be harnessed by the detectors for a source at the same sky position. For observation times exactly multiple of 1 yr or much longer the difference between the average optimal signal-to-noise ratio in these detectors and ground-based ones is easily accounted for with a simple rescaling by a factor of $\sqrt{3/4}$ of the amplitude spectral density of the space-based detectors as described in Appendix \ref{sec:appendix}.
\item The (quadratic) signal-to-noise ratio scales linearly with the the number of identical detectors $N$ contributing coincident data to the search \citep{2005PhRvD..72f3006C}.
This trivially translates in terms of sensitivity depth as follows 
\begin{equation}
\label{eq:depth_dependence}
\mathcal{D} = \frac{\sqrt{S_h(f)}}{h_0} \sqrt{N} ,
\end{equation}
where we omit the confidence interval superscript for ease of notation. In the ground-based detector searches that we consider, data is combined from the two LIGO detectors.

\citet{2011PhRvD..83d4011Y} show that both DECIGO and Big Bang Observer will effectively constitute an observatory of eight identical detectors; we therefore estimate the smallest signal detectable by DECIGO/BBO detectors to be
\begin{equation}
\label{eq:h0Decigo}
h_0^{\textrm{DEC/BBO}}(f)= \frac{\sqrt{S_h^{\textrm{DEC/BBO}}(f)}}{\mathcal{D}_\textrm{LIGO}}\sqrt{\frac{2}{8}}.
\end{equation}

The Einstein Telescope consists of 3 independent identical detectors \citep{2011CQGra..28i4013H}.  
For Cosmic Explorer a 40 km-arm detector is planned plus a less sensitive, 20 km arm, one. When the sensitivities are factors of two or more different, the gain for continuous wave searches is limited, and comes with the extra cost of processing the other data stream, so data from the least sensitive detectors is typically not considered. So for Cosmic Explorer we assume a single detector. 
Having made these considerations, we estimate the smallest signal detectable by Einstein Telescope to be
\begin{equation}
\label{eq:h0ET}
h_0^{\textrm{ET}}(f)= \frac{\sqrt{S_h^{\textrm{ET}}(f)}}{\mathcal{D}_\textrm{LIGO}}\sqrt{\frac{2}{3}},
\end{equation}
while for Cosmic Explorer
\begin{equation}
\label{eq:h0CE}
h_0^{\textrm{CE}}(f)= \frac{\sqrt{S_h^{\textrm{CE}}(f)}}{\mathcal{D}_\textrm{LIGO}}\sqrt{2}.
\end{equation}
\end{itemize}

\section{Continuous waves from highly magnetized pulsars}
\label{sec:known_pulsars}
We start our analysis considering known pulsars. We assume that they possess a permanent ellipticity, that sources a continuous stream of gravitational waves. Indirect constraints on the ellipticity of a pulsar can be derived starting from its timing measurements, by equating the entire observed loss of rotational kinetic energy to the energy loss due to continuous wave emission. The ellipticity that is necessary to support emission at this level is the so-called spin-down limit ellipticity:
\begin{equation}
\label{eq:eps_sd}
\varepsilon^{sd} = \sqrt{\frac{5 c^5}{512 \pi^4 G I}\frac{|\dot{\nu}|}{\nu^5}}, 
\end{equation}
where $\nu$ is the spin frequency. Cast in terms of the gravitational wave intrinsic amplitude for a source at a distance $D$, Eq.~\ref{eq:eps_sd} reads
\begin{equation}
\label{eq:h0_sd}
h_0^{sd} =  \frac{1}{D} \sqrt{\frac{5 G I |\dot{\nu}|}{2 c^3 \nu}}. 
\end{equation}
The spin-down limit constitutes a strong indirect observational constraint on the maximum possible non-axisymmetry given the observed rate at which the pulsar looses rotational kinetic energy. In fact, since a considerable fraction of rotational power is emitted in the electromagnetic spectrum, it is realistically expected that pulsars have ellipticities significantly smaller than their spin-down upper limit. LIGO observations have constrained the ellipticity of two dozens pulsars to be smaller than the respective spin-down upper limit, this constraint being the tightest for the Crab pulsar, whose emission amplitude is about a factor of 100 below the spin-down limit \citep{2022ApJ...935....1A,2025arXiv250101495T}.

We consider highly magnetised neutron stars, assume that they are misaligned rotators and consider continuous gravitational wave emission sourced by the magnetic deformation. We look for pulsars that could be such objects. 
We select pulsars from the ATNF pulsar catalogue \citep{Manchester_2005} and from the McGill magnetar catalog \citep{2014ApJS..212....6O} having surface magnetic field 
\begin{equation}
\label{eq:Bselection}
B_{s} \geq B_{s}^{crab} = 3.8 \cdot 10^{12} \, \textrm{G} .
\end{equation}

Since we imagine carrying out a targeted search over years, we keep only the pulsars with known frequency and frequency derivatives that we can use to define the target waveform. We consider only the $f=2\nu$ harmonic here. 
We exclude pulsars that present significant glitching activity, that we define as having a Poisson probability larger than 50\% to glitch during a 2 year observation, based on the glitch rate of the glitch catalogues considered in Section \ref{sec:glitch_source}. We also exclude pulsars whose $\dot{\nu}$ is not negative, since for those we would not be able to estimate the spin-down limit.
We find 300 highly magnetised neutron stars that fit all these constraints and are hence suitable for a fully coherent targeted search.

We imagine repeating over each of these pulsars the search of \citet{2021ApJ...923...85A} using data from future detectors. \cite{2021ApJ...923...85A} combine LIGO Hanford and Livingston data approximately coincident over about two years, and target seven different pulsars.
The average sensitivity depth attained over the seven targets is $\approx 750 ~[\textrm{Hz}]^{-1/2}$. 

We estimate the smallest detectable signal from each of these pulsars by DECIGO/BBO, Einstein Telescope and Cosmic Explorer using Eq.s  \ref{eq:h0Decigo}, \ref{eq:h0ET} and \ref{eq:h0CE} with $\mathcal{D}_\textrm{LIGO} \approx 750 ~[\textrm{Hz}]^{-1/2}$ and obtain

\begin{equation}
\label{eq:detectableh0}
\begin{split}
&h_0^{\textrm{DEC/BBO}}(f)= \frac{\sqrt{S_h^{\textrm{DEC/BBO}}(f)}}{1500 ~[\textrm{Hz}]^{-1/2}}\\
&h_0^{\textrm{ET}}(f)= \frac{\sqrt{S_h^{\textrm{ET}}(f)}}{918 ~[\textrm{Hz}]^{-1/2}}\\
&h_0^{\textrm{CE}}(f)= \frac{\sqrt{S_h^{\textrm{CE}}(f)}}{530 ~[\textrm{Hz}]^{-1/2}}.
\end{split}
\end{equation}

We consider potentially detectable any pulsar for which the smallest detectable amplitude from Eq.s \ref{eq:detectableh0} is smaller than the spindown limit $h_0^{sd}$ from Eq.~\ref{eq:h0_sd}. 

\subsection{Results}
\label{sec:permanent_results}
Because of the very similar sensitivity and frequency coverage, results from DECIGO and Big Bang Observer are grouped together, as well as those from Einstein Telescope and Cosmic Explorer.

We convert smallest detectable amplitudes into smallest detectable ellipticities, using Eq. \ref{eq:h0}, with $I=1.4\times 10^{38}$ kg m$^2$, corresponding to  $I = 0.35 \cdot M R^2$, valid for a neutron star described by a polytrope with polytropic index $n=1$ \citep{10.1111/j.1365-2966.2008.14054.x}, for $M=1.4 M_{\odot}$ and $R=12 ~\textrm{km}$.

We consider the magnetar 4U 0142+61 for which \cite{2014PhRvL.112q1102M} have inferred an ellipticity of $\approx 1.6 \cdot 10^{-4}$, based on an observed x-ray pulse periodicity consistent with free precession. 
We use this as an ellipticity reference value for what ellipticity highly magnetised neutron stars may present, and in Table \ref{tab:permanent_known_sources} we show how many detectable sources would present an ellipticity smaller/equal to $1.6 \cdot 10^{-4}$, assuming an amplitude at exactly the level of the smallest detectable (Eq.s~\ref{eq:detectableh0}). 
As detailed below, an ellipticity of order $10^{-4}$ is large and has a place in models of magnetic deformation (see for example \citet{2023ApJ...955...19D}). Models that account for elastic crust support, generally predict ellipticities $\lesssim 10^{-5}$ \citep{2000MNRAS.319..902U, 2006MNRAS.373.1423H, 2009PhRvL.102s1102H, 2021MNRAS.500.5570G, 2022MNRAS.517.5610M}.

Deci-Hz space-based detectors have the highest count of highly magnetised pulsars with smallest detectable ellipticity below this reference ellipticity: 35 sources versus 29 for ET/CE. We point out however, that 20 of the 35 deci-Hz sources are also detectable by ET with a smaller ellipticity.
\begin{table}
	\centering
	\caption{Number of pulsars emitting continuous gravitational waves sources potentially detectable by various detectors, i.e. with detectable $h_0$ from Eq.s~\ref{eq:detectableh0} smaller/equal to $h_0^{sd}$. For the DECIGO/BBO and ET/CE we have considered the sample 300 highly magnetised pulsars and magnetars discussed in Section \ref{sec:known_pulsars}. 
The LIGO column refers to \citet{2022ApJ...935....1A}, therefore the percentages shown are referred to the 255 pulsars targeted in that study.}
	\label{tab:permanent_known_sources}
	\begin{tabular}{cccccc}
		\toprule
		\textbf{DECIGO/BBO}& &\textbf{ET/CE}& & &\textbf{LIGO}\\
		\midrule
		\multicolumn{6}{c}{$h_0 \leq h_0^{sd}$}\\
		&&&&\\
		\multicolumn{1}{c}{\textbf{300} ($100\,\%$) }& &\multicolumn{1}{c}{\textbf{82} ($27 \,\%$) }& & &\multicolumn{1}{c}{\textbf{29} ($11.4\,\%$)}\\
		\cmidrule(lr){1-6}
		\multicolumn{6}{c}{$\frac{h_0}{h_0^{sd}} \! < \! 0.1$}\\
		&&&&\\
		\multicolumn{1}{c}{\textbf{284} ($95 \,\%$) } & & \multicolumn{1}{c}{\textbf{47} ($16 \,\%$) } & & & \multicolumn{1}{c}{\textbf{5} ($2\,\%$) }\\
		\cmidrule(lr){1-6}
		\multicolumn{6}{c}{$h_0 \! \leq \! h_0^{sd} \, ~{\textrm{and}}~\, \varepsilon \! < \! 1.6 \cdot \! 10^{-4}$}\\
		&&&&\\
		\multicolumn{1}{c}{\textbf{35} ($12 \,\%$) } & & \multicolumn{1}{c}{\textbf{29} ($10 \,\%$) } & & & \multicolumn{1}{c}{\textbf{17} ($6.7\,\%$)}\\
		\bottomrule
	\end{tabular}
\end{table}
\begin{figure*}
\caption{There are 300 coloured markers (stars and crosses), corresponding to the smallest detectable ellipticities of sources from the considered samples. Different colours refer to detectability by different detectors as specified in the legend. If a source is detectable by both DECIGO/BBO and ET/CE, we plot only the smallest between the detectable ellipticities, using the colour that refers to the detector capable to detect it. 
The grey triangles show the spin-down ellipticity for every source. The horizontal dotted line is the ellipticity inferred by \citet{2014PhRvL.112q1102M} for the magnetar 4U 0142+61. The dash-dotted lines indicate characteristic ages assuming only gravitational wave braking.}
\includegraphics[width=\textwidth]{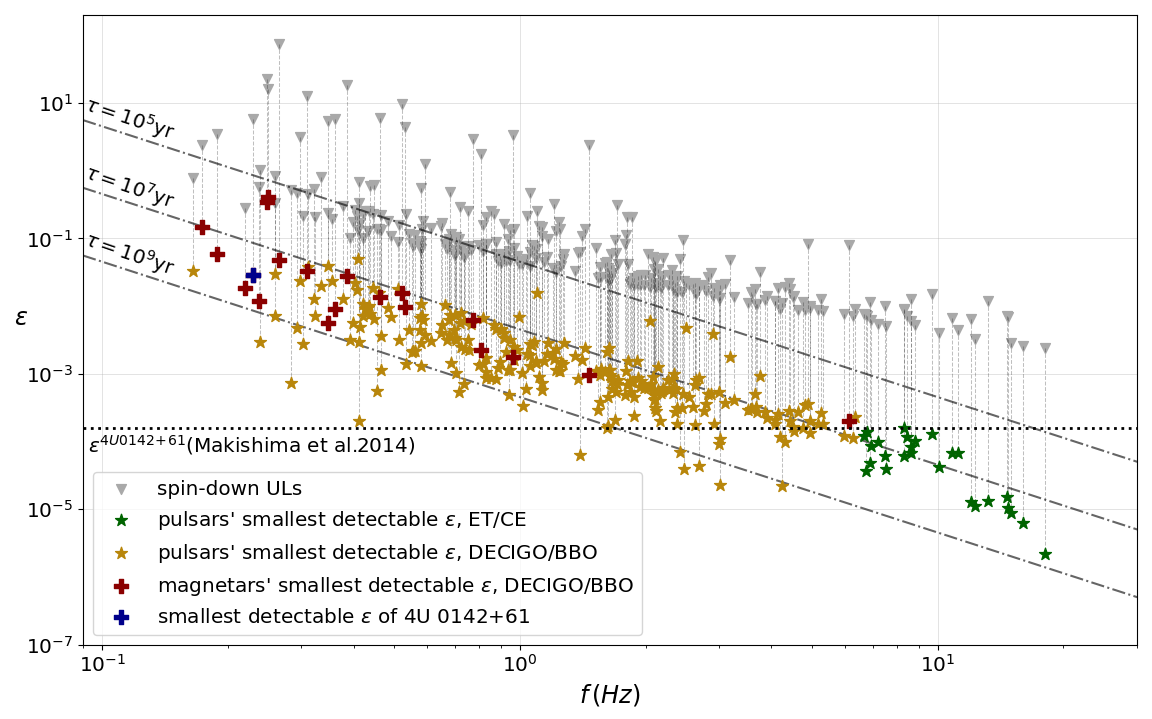}
\label{fig:ellipticity_constraints}
\end{figure*}

Just in terms of detectability, i.e.  beating the spin-down limit, all of the considered sources are ``detectable" by DECIGO/BBO while less than 30\% are detectable by ET/CE; and for $\approx$ 95\%, over 280 sources in the sample, the DECIGO/BBO detectability level lies more than a factor of 10 {\it{below}} the spin-down limit, whereas for ET/CE there are less than 50.

Figure~\ref{fig:ellipticity_constraints} shows the spin-down limit ellipticities for all potentially detectable sources in the considered samples and the expected detectability level at each pulsar frequency. We remind the reader that the detectability condition requires that the smallest detectable $h_0$ from Eq.s~\ref{eq:detectableh0} be smaller/equal than $h_0^{sd}$. Our results are broadly consistent with those of \cite{miller2025searchingcontinuousgravitationalwaves}, albeit on a different pulsar sample that also includes low/mildly magnetised neutron stars spinning below 10 Hz. We excluded these stars because we did not want to invoke an unspecified deformation mechanism and they cannot be sufficiently deformed by the magnetic field to be detectable.

The horizontal dotted line corresponds to the inferred ellipticity of the magnetar 4U 0142+61. Unfortunately the smallest detectable ellipticity for this same magnetar (highlighted in blue) is about 2 orders of magnitude larger than the value inferred by \citet{2014PhRvL.112q1102M}. However, a total of 42 sources are potentially detectable with DECIGO/BBO and 3rd generation detectors with an ellipticity smaller than this reference value.

The dash-dotted lines represent characteristic ages $\tau$ of sources as a function of frequency and ellipticity, assuming that they spin-down only due to gravitational wave emission:
\begin{equation}
\label{eq:char_age}
\tau = \frac{5c^2}{2048\pi^4 G}\frac{1}{I\varepsilon^2 \nu^4}.
\end{equation}
Figure \ref{fig:ellipticity_constraints} shows that no star or cross lies above the $10^5 \, \textrm{yr}$ age line, with most of them being below the $10^7 \, \textrm{yr}$ line. This indicates that our potentially detectable sources do not need to be unrealistically young in order to have a deformation at the detectable level.

\begin{figure}
\caption{Cartoon showing the effect that different poloidal-toroidal mixtures have on the shape of a neutron star. A toroidal magnetic field (left) makes the star prolate, whereas a poloidal field makes it oblate. The dotted axis is the rotation axis, inclined by an angle $\alpha$ with respect to the axis of symmetry of the magnetic field (solid/dashed line).}
\includegraphics[width=\columnwidth]{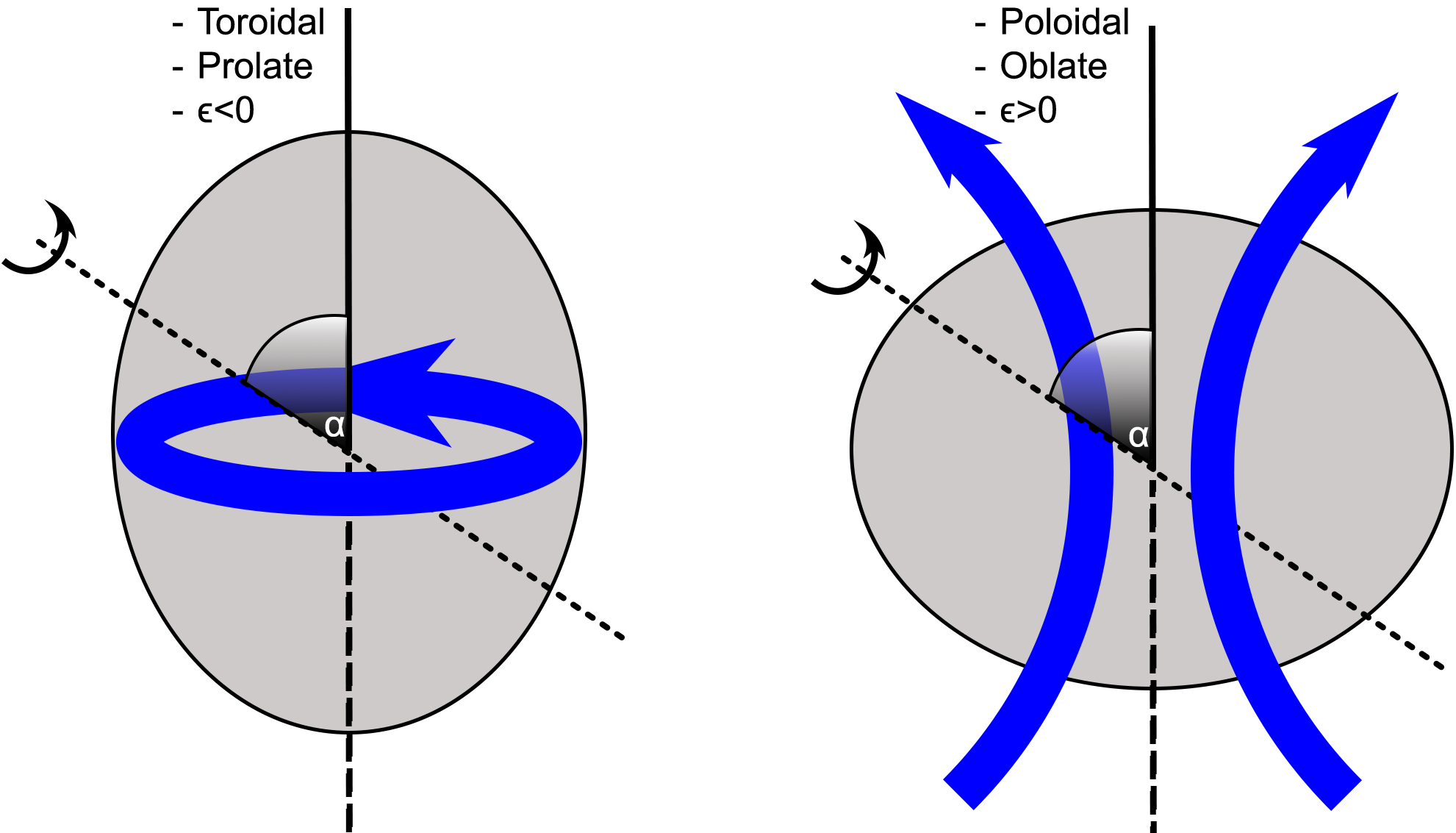}
\label{fig:prolate_oblate}
\end{figure}

Very little to nothing is known about the internal magnetic field of neutron stars. 
This means that if a feature of the star sourced by the internal magnetic field -- like the ellipticity --  can give rise to a measurable effect -- like a continuous gravitational wave -- it has the potential to provide insight into an otherwise obscure quantity. 

For instance, \cite{2011MNRAS.417.2288M} propose a model to calculate the magnetic deformation in neutron stars. They assume an internal poloidal-toroidal magnetic field mixture matched with an external dipole field. An interesting feature of their model is that it allows to calculate the ellipticity from the observed value of the surface magnetic field $B_s$. This is made possible by a continuity constraint between the internal mixed field and the external dipole. This effectively transfers the internal field dependency to a free parameter, $\Lambda$, defined as
\begin{equation}
\label{eq:lambda_definition}
\Lambda = \frac{\int_{V} dV \, \mathbf{B_p}^2}{\int_{V} dV \, \mathbf{B}^2} ,
\end{equation}
corresponding to the fraction of magnetic energy stored in the poloidal component ($\mathbf{B_p}$ above) over the total magnetic energy of the neutron star, hence a value between 0 and 1. In Eq. \ref{eq:lambda_definition}, the integrals are performed over the volume $V$ occupied by the different magnetic field components. 

The ellipticity they obtain in the case of a neutron star described by a polytrope with $n=1$, considering the geometrical simplification of an orthogonal rotator ($\alpha = \pi/2$ in Eq. \ref{eq:ellipticityVsB}), is
\begin{equation}
\label{eq:mastrano}
\varepsilon \! = \! \epsilon_0  \frac{\Lambda-\Lambda^\star}{\Lambda}, 
\end{equation}
with 
\begin{equation}
\begin{split}
& \epsilon_0 = 5.2\times 10^{-5} \left( \frac{B_s}{10^{15} \textrm{G}} \right)^{\!2} \! \left( \frac{1.4\textrm{M}_{\odot}}{M} \right)^{\!2} \! \left( \frac{R}{12\textrm{km}} \right)^{\!4} ,\\
& \Lambda^\star = 0.385 .
\end{split}
\label{eq:epsilon0}
\end{equation}
The ellipticity vanishes when $\Lambda=\Lambda^\star$. For $\Lambda < \Lambda^\star$ the ellipticity $\varepsilon <0$ and the star is prolate. For $\Lambda > \Lambda^\star$ the ellipticity $\varepsilon > 0$ and the star is oblate (see Fig.~\ref{fig:prolate_oblate}). 

In the case of a non-detection, the upper limits on the ellipticity $\epsilon_{UL}$ of a known pulsar, can be used to constrain the poloidal-to-total magnetic field energy ratio $\Lambda$ as follows
\begin{equation}
\label{eq:LambdaConstrain}
    \begin{cases}
      \Lambda > \Lambda^\star \frac{\epsilon_0}{\epsilon_0 + \epsilon_{UL}} & \textrm{if}~\epsilon_{UL} \geq \epsilon_0\\
      \Lambda^\star \frac{\epsilon_0}{\epsilon_0 + \epsilon_{UL}} < 
      \Lambda < \min (\Lambda^\star \frac{\epsilon_0}{\epsilon_0 - \epsilon_{UL}}, 1) & \textrm{if}~\epsilon_{UL} < \epsilon_0 .
    \end{cases}       
\end{equation} 

\begin{figure}
\caption{Ellipticity as a function of $\Lambda$ for the generic mixed toroidal-poloidal field model of Eq.~\ref{eq:mastrano}, with $\epsilon_0$ and $\Lambda^\star$ from Eq.s~\ref{eq:epsilon0} with the magnetic field of J1513-5108 (lower red crosses). For this model yields a prolate or oblate object depending on whether $\Lambda\gtrless 0.385$. We also show the same curve for the specific model constructed for SGR 0418, defined by Eq.s~\ref{eq:mastrano}, \ref{eq:mastrano_2} (upper blue crosses). With a $\Lambda^\star\approx 0.99$, the model favours prolate configurations. The straight line are the expected smallest detectable signals (or upper limits) from searches for emission from these objects with deci-Hz detectors.}
\includegraphics[width=\columnwidth]{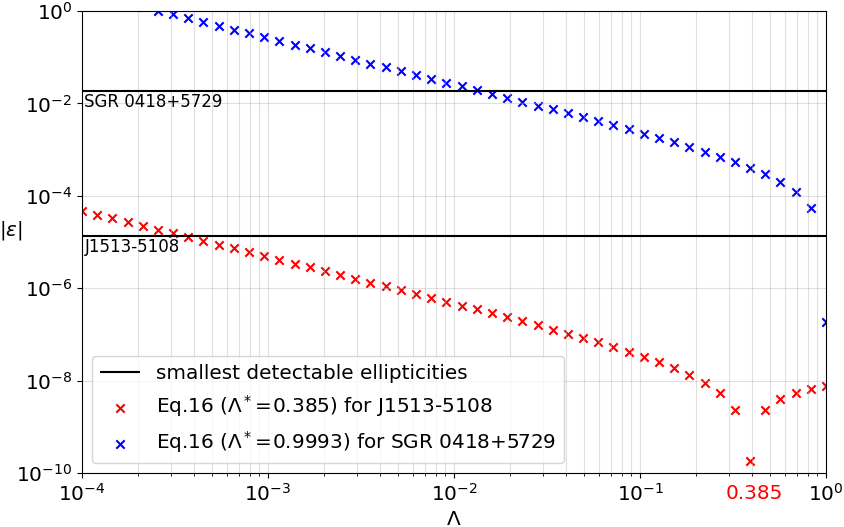}
\label{fig:EpsilonVersusLambda}
\end{figure}

Let's consider a concrete example. The surface magnetic field of J1513-5108, inferred from the observed spin-down, is $\approx 1.5\cdot10^{13}\,\textrm{G}$. Plugging these two values in Eq. \ref{eq:epsilon0} we find $\epsilon_0=1.2\times 10^{-8}$ for $M=1.4M_{\odot}$, $R=12 \textrm{km}$. The smallest detectable signal (or upper limit) from a deci-Hz continuous wave search is expected to be at  $\approx 1.3 \cdot 10^{-5}$ so a non-detection would constrain $\Lambda > 3.6\cdot10^{-4}$, as one can see from the lower curve in Fig.~\ref{fig:EpsilonVersusLambda}.

\subsubsection{SGR 0418+5729}
\label{sec:sgr0418}
This magnetar has a relatively low magnetic field of $B \! \approx \! 6 \cdot 10^{12} \, \textrm{G}$, inferred from its spin-down rate.
Observations in the X-ray spectrum provide independent magnetic field measurements, since in general spectral features in this band depend on the value of the magnetic field near the surface of the star.

\citet{2011MNRAS.418.2773G} and \citet{2013Natur.500..312T} independently arrive to the conclusion that the surface magnetic field in SGR 0418+5729 must be of order $B_s \approx 10^{14} \, \textrm{G}$ and that small scale multipolar magnetic field components, which don't play a role in the observed spin-down of the magnetar, might be the cause of the discrepancy between this value and the one inferred from timing measurements.
An alternative interpretation of this discrepancy is given by \citet{2021ApJ...913L..12M}, who attributes it to SGR 0418+5729 being an ``almost-aligned" rotator in a plasma-filled magnetosphere.

\citet{2015MNRAS.447.3475M} generalise their previous magnetic deformation model accounting for the presence of multipolar components within the stars' surface.
As a worked example they apply their model to SGR 0418+5729, assuming the usual poloidal-toroidal magnetic field decomposition, but expanding the poloidal component as a sum of a dipolar and a quadrupolar component.
The ellipticity is again given by Eq. \ref{eq:mastrano}, but this time with:
\begin{equation}
\begin{split}
&\epsilon_0 \! = \! 2.6 \times 10^{-2} \left( \frac{B_s}{10^{15} \textrm{G}} \right)^{\!2} \! \left(\frac{1.4\textrm{M}_{\odot}}{M} \right)^{\!2} \! \left(\frac{R}{12\textrm{km}} \right)^{\!4} ,\\
&\Lambda^\star = 0.9993 .
\end{split}
\label{eq:mastrano_2}
\end{equation}
We emphasise that the Eq.s \ref{eq:mastrano_2} are not general, and apply to SGR 0418+5729 only.
The comparison between Eq.s \ref{eq:epsilon0} and \ref{eq:mastrano_2} suggests that the presence of a quadrupolar component makes the star ``more prolate" in comparison with the purely dipolar case.

Following the same sort of reasoning as done for J1513-5108 at the end of the previous section, with an expected minimum detectable ellipticity of $\approx 1.8 \cdot 10^{-2}$, a null result from SGR 0418+5729 would constrain $\Lambda > 1.4\cdot10^{-2}$ -- see upper curve in Fig.~\ref{fig:EpsilonVersusLambda}.

In principle, from a constraint on $\Lambda$ we can derive a constraint on the internal magnetic field through Eq. \ref{eq:lambda_definition}. In reality though, the relative volumes occupied by the two magnetic field components remain unknown, therefore such constraints can only give an order-of-magnitude indication. 
This also highlights the importance of having additional observational data in the electromagnetic spectrum (like X-ray observations for SGR 0418+5729), as well as neutron-star modeling studies to help refine the understanding of neutron stars and prepare to interpret gravitational wave observations.

The discussion here should be understood as an example of the kind of science that one can do with continuous gravitational wave search results.

\section{Continuous waves from central compact objects in supernova remnants}
\label{sec:cco_snr}
A conventional target for continuous waves searches are central compact objects in supernova remnants. These are X-ray sources near the centre of supernova remnants, tipically associated with young and hot neutron stars. 

If we detect pulsations from the central compact object, these targets are no different than known pulsars from the point of view of a search.
In the absence of pulsations instead, the sky position is known but the rest of phase parameters are not, so we need to set-up a grid that covers the parameter space of interest. This is generally done considering the kinematic age of the supernova remnant as a general guideline \citep{2018PhRvD..97b4051M}. The resulting high computational cost of this type of searches (\emph{directed searches}) translates into worse sensitivity compared to the targeted case.

The reasons why central compact objects in supernova remnants are interesting targets for continuous waves emission are mainly two and related with their young age:
firstly, supernova explosions and collapse are not perfectly axi-symmetric processes hence deformations from the birth of the neutron star are plausible. These might ``wash out" as the neutron star ages, so younger objects have more chances to possess high ellipticities;
secondly, young/newborn neutron stars should have very high spin rates and as discussed already, with equal deformations, fast rotators are stronger emitters.

At least for the class of central compact objects, the second point is in tension with the observational evidence that of the 10 known ones, 3 show pulsations which correspond to relatively slow rotators (the fastest is spinning at $\approx 9.5 \, \textrm{Hz}$).
A peculiar characteristic of pulsating central compact objects is a surprisingly low surface magnetic field (considering they are non-recycled young neutron stars, $B \approx 10^{10-11} \, \textrm{G}$) sometimes combined with magnetar-like activity, which instead require strong magnetic fields to power it.

One possible explanation for central compact objects' phenomenology is that these are young neutron stars whose magnetic field has been buried by fallback accretion \citep{2007ApJ...665.1304H}, phenomenon that may also be the cause for slow rotation due to propeller effect which spins-down the star \citep{2022ApJ...934..184R}.

These buried fields can reach magnetar-like regimes, thus power X-ray bursts.
This is probably the case of the enigmatically slow rotator 1E 161348–5055 in the supernova remnant RCW 103 \citep{2006Sci...313..814D, 2016ApJ...828L..13R} for which however one cannot estimate the magnetic field due to the lack of spin-down measurements.
There are other, less clear examples, where there is circumstantial evidence of magnetar-like features in supernova remnants hosting a central compact object.

PSR J1852+0040 at the centre of Kesteven 79 for example, presents highly non-uniform surface temperature likely linked to the presence of a much stronger magnetic field than the one inferred from X-ray timing \citep{2014ApJ...790...94B}.
The frequently targeted central compact object in Cassiopeia A supernova remnant also shows peculiarities. \citet{2005Sci...308.1604K} identify close to speed of light moving features within the supernova remnant region, which they interpret as the result of the echoed re-emission of interstellar dust around the remnant heated up by outward moving photons. The detected phenomenon is consistent with the energetics of giant flares in magnetars. 

The above-mentioned objects can be cases of \emph{hidden magnetars} \citep{1999AA...345..847G}, that is neutron stars with an enormous magnetic field temporarily concealed by the effect of fallback accretion burial. Such hypothetical objects might be slow rotators (just as known pulsating central compact objects) and might possess substantial deformations due to the strong magnetic fields buried within the surface. 

We want here to briefly outline future prospects for the detection of continuous waves emitted by hidden magnetars.
We consider both J1852+0040, referred to by \citet{2015PASA...32...18P} as the most prominent hidden magnetar candidate, and the central compact object in Cassiopeia A.

\subsection{PSR J1852+0040}
\label{sec:J1852}
For a search targeting PSR J1852+0040 we assume to have a timing solution and hence the smallest detectable signal amplitudes by various detectors to be those of Eq.s~\ref{eq:h0Decigo},\ref{eq:h0ET} and \ref{eq:h0CE}.

We find PSR J1852+0040 to be detectable by Cosmic Explorer if it is emitting at least $\approx 20\%$ of its spin-down luminosity in continuous waves or equivalently if its ellipticity is $\gtrsim 7.2\cdot10^{-6}$. This ellipticity is consistent with an internal field of the order of $10^{15}\,\textrm{G}$ \citep{2008MNRAS.385..531H}, value perfectly in line with this source being an hidden magnetar (see Figure~ \ref{fig:hidden_magnetar}).

\subsection{Cassiopeia A}
\label{sec:CasA}
In a search for continuous gravitational waves from a supernova remnant with no detected pulsations, we define the parameter space region to cover, following \citet{Ming2016}: 
\begin{equation}
    \begin{cases}
        &-f / (n-1) \tau \ \leq \dot{f} \leq \ 0\\
        &0 \ \leq \ddot{f} \leq \ n f/{\tau}^2 .
    \end{cases}
\label{eq:param_space}
\end{equation}
In the above Equations, $\tau$ is the age of the remnant and $n$ is the neutron star's breaking index.
To make our calculations explicit, we consider the supernova remnant Cassiopeia A, which has an estimated age of 330 years \citep{Fesen2006}.
We also generously assume the breaking index to be equal to 2 in the first equation and to 7 in the second, encompassing the widest $\dot{f},\ddot{f}$ ranges \citep{Morales:2025skm,Ming:2024dug}.
We consider two different frequency ranges for the two groups of detectors defined in Section \ref{sec:permanent_results}.

For DECIGO and Big Bang Observer we consider the range 
\begin{equation}
\label{eq:CasAfRangeDecigo}
0~\textrm{Hz} \leq f  \leq 7~\textrm{Hz} .
\end{equation}
This choice accounts for the fact that 3rd generation and deci-Hz detectors have comparable sensitivities around 5 Hz, and as we go towards higher frequencies, the former definitely outperforms the latter for the rest of the band.

Next we estimate the search sensitivity. We assume a total observation time of 4 years and consider a hierarchical search comprising searches with the same coherence length and search set-ups. We compare of order a thousand different search set-ups, corresponding to different coherent lengths and grid spacings. We choose the one that yields the highest sensitivity and is doable within the computational budget limit. We set the latter to be $\approx 50,000$ CPU core years which is roughly the computational power of \emph{Einstein@home} \citep{Anderson2004,Anderson2006} running for one year.
\begin{table}
	\centering
	\caption{Optimal search setups and their estimated $\mathcal{D}^{90 \%}$ for a hierarchy of 4 fully coherent search stages, each with coherence time $T_{coh}=1\textrm{yr}$. The final false alarm is set at $<1\%$. The ditto mark means ``same as above".}
	\label{tab:search_setup}
	\begin{tabular}{||c|c|c|c||}
		\toprule
		Detectors&$f$ range [Hz]&no. of templates&$\mathcal{D}^{90 \%}$\\
		\hline
		\hline
		DECIGO/BBO&0-7&$2.3 \times 10^{16}$&599\\
		Einstein Telescope&7-30&$1.8\times 10^{18}$&351\\
		Cosmic Explorer&"&"&203\\
		\bottomrule	
	\end{tabular}
\end{table}
The optimal set-up uses a small fraction of the computing budget, has a coherence length $T_{coh} = 1 \, \textrm{yr}$, covers the entire parameter space with $\approx 2 \cdot 10^{16}$ templates and has of a total of 3 follow-up stages on independent data streams. 

We find this search to be able to attain a sensitivity depth of $\mathcal{D}^{90\%} \approx 600~[\textrm{Hz}]^{-1/2}$ with a false alarm rate of $<1\%$  for the entire hierarchical search\footnote{We mean that in noise we expect to have to perform more than 100 hierarchical searches in order for one of them to yield a candidate at the level of $\mathcal{D}^{90\%} \approx 600~[\textrm{Hz}]^{-1/2}$.} (see Table \ref{tab:search_setup}).

For 3rd generation ground-based detectors Einstein Telescope and Cosmic Explorer we consider the range 
\begin{equation}
\label{eq:CasAfRangeCE}
7~\textrm{Hz} \leq f  \leq 30~\textrm{Hz} .
\end{equation}
The most sensitive search possible consists of a fully coherent first stage where we cover the entire parameter space with $\approx 2 \cdot 10^{18}$ templates with a coherence time of $T_{coh} = 1 \, \textrm{yr}$, again corroborated by 3 follow-up stages.
We obtain different sensitivity depths for Einstein Telescope and Cosmic Explorer due to the different number of effective detectors.
In comparison with the DECIGO/BBO searches, these are less sensitive because of the much bigger parameter space to cover. 

\begin{figure}
\caption{The blue scatter points are currently the best ellipticity constraints for the central compact object in Cas A at these frequencies \citep{2024PhRvD.110d2006W}. The upper dashed lines are the estimated smallest detectable ellipticities by DECIGO/BBO (at lower frequencies) and CE (at higher frequencies) for Cas A. We also consider putative sources, with a similar age to Cas A, but at smaller distances (Cas A is at about 3.4 kpc). The lower dashed curves show the smallest detectable ellipticities for these. The single markers refer to the pulsar J1852+0040 in Kes 76.}
\includegraphics[width=\columnwidth]{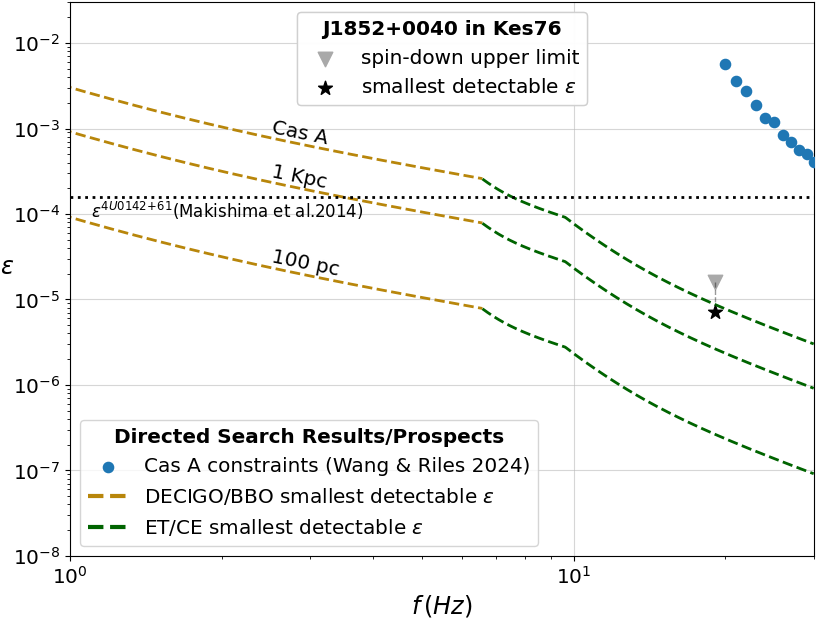}
\label{fig:hidden_magnetar}
\end{figure}

The suggested searches outperform current results at $\approx 20$ Hz by nearly three orders of magnitude (see Fig.~\ref{fig:hidden_magnetar}). This is not only due to the improved 3rd generation detectors' sensitivity but also to the search itself that can be performed at these low frequencies. We see that such very sensitive search is needed if one wants to probe ellipticities below the reference value of $1.6 \cdot 10^{-4}$ for sources at a few kpcs. 

For a putative Cas A-like source at 100 pc we see that the smallest detectable ellipticities are accordingly much smaller, and in fact stay below the reference value also in the DECIGO/BBO band. We must note however that if we consider the synthetic population introduced in our previous work \citep{2023ApJ...952..123P} to be representative of our Galaxy, we find no sources within 100 pc to be younger than $10^5 \, \textrm{yr}$. 

Vela Jr. is a promising young supernova remnant, for which the search described here could meaningfully apply. Since it is at $\approx 1 \textrm{kpc}$, the smallest detectable ellipticity is smaller than for Cas A. Nonetheless Vela Jr. does not present any magnetar-like activity to date.

\section{Glitching pulsars}
\label{sec:glitch_source}
Over two hundred pulsars are known to ``glitch", i.e. their secular spin frequency evolution is occasionally interrupted by a sudden spin-up (an increase in spin frequency). 
The complex and varied phenomenology of glitching pulsars is not completely understood. 
The model most commonly adopted to explain it involves the presence of two fluids in the star: a superfluid component, well below the crust, and a normal component that includes the inner crust and the crust itself. 
As the neutron star spins-down, the angular velocity difference between the normal component, which follows the observable spin-evolution, and the superfluid one, which instead spins decoupled from the latter, will reach a critical value so that part of the excess angular momentum of the superfluid component is \emph{suddenly} transferred to the crust, thus the sudden increase in its angular velocity. There are several proposed mechanisms through which this angular momentum transfer can happen \citep{1991ApJ...382..587R, 1991ApJ...373..592L, 2000A&A...361..795C, 2003PhRvL..90i1101A, 2022ApJ...941..148L}. 

In the aftermath of a glitch, a recovery -- partial or total -- of the spin to the pre-glitch state is observed. In general the observed jump in the pulsar's spin frequency $\delta \nu$ can be expressed as the sum of a permanent component $\delta \nu_p$ and a transient one $\delta \nu_t$.
If the recovery is total $\delta \nu_p = 0$, the frequency change is purely transient and, after some time (typically of the order of days) the spin state of the pulsar is back to the pre-glitch value. Conversely, if $\delta \nu_p \neq 0$ the recovery is partial and the star will have gained some permanent residual spin (and possibly also spin time derivatives).

The \emph{healing parameter}
\begin{equation}
\label{eq:heal_param}
\mathcal{Q} \coloneq 1- \frac{\delta \nu_p}{\delta \nu}
\end{equation}
ranges between 0 and 1. If $\mathcal{Q} \! = \! 0$ the pulsar frequency does not recover from the glitch, and all the frequency increase due to the glitch is permanent. When $\mathcal{Q} \! = \! 1$ the pre-glitch frequency is fully restored aver a recovery period (see \citet{2020MNRAS.498.3138Y} for a more in-depth discussion). 
$\mathcal{Q}$ can be measured, although this is not easily done as it requires extended observations after the glitch event.

Continuous waves might play a role have in restoring the spin frequency of a pulsar after a glitch.
We will assume that the excess energy coming from the super-fluid component rotating faster than the crust just before the glitch, is dissipated by the emission of gravitational waves during a time $\tau$ due to a quadrupolar deformation formed at the time of the glitch when angular momentum was suddenly transferred from the superfluid to the crust.  
The gravitational wave amplitude that supports emission at this level is \citep{2011PhRvD..84b3007P}
\begin{equation}
\label{eq:h0_sd_glitch}
h_{0, r}^{sd} = \frac{1}{d}\sqrt{\mathcal{Q} \, \frac{5 G I}{2 c^3} \frac{1}{\tau} \frac{\delta \nu}{\nu}},
\end{equation}
and it is referred to as the \emph{superfluid upper limit}.
\begin{figure}
\caption{Glitch amplitudes for the considered sample. The x-axis is neutron star spin frequency $\nu$.
Glitches that have a concurrent measurement of the healing parameter are indicated with thick markers.}
\includegraphics[width=\columnwidth]{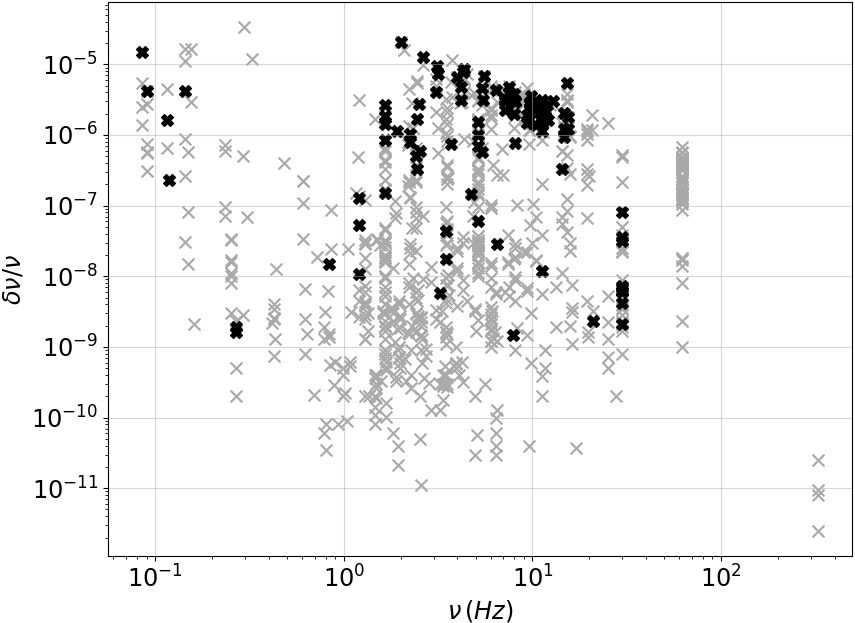}
\label{fig:glitch_distribution}
\end{figure}

We consider glitch data from the ATNF \citep{Manchester_2005} and the Jodrell bank \citep{2011MNRAS.414.1679E} glitch catalogues.
At the time of writing, the ATNF glitch catalogue contains 625 entries from 211 sources while the Jodrell bank catalogue contains 679 entries from 226 sources. The two catalogues have significant overlap. We proceed to merge them, following the same criteria for recognising a duplicate as \cite{2023MNRAS.519.5161M}, that is assuming a glitch to be the same one if associated with the same source and if the occurrence time of the entries differ from each other by less than one day.
We then match the glitches with known pulsars or magnetars from either the ATNF pulsar catalogue or the McGill magnetar catalogue \citep{2014ApJS..212....6O}, considering only pulsars or magnetars with known distances. We prioritise again entries from the McGill catalogue over the ATNF ones for the case of magnetars.
We also remove the two anti-glitches present in the merged catalogue since they do not apply to the scenario considered here. Lastly, we note that two glitches with the exact same parameters are associated with the same pulsar under different names (J1636-2614/J1636-2626) in the two catalogues. We treat these as a duplicate.

We identify 789 glitches from 216 pulsars and magnetars, of which 115 from 58 pulsars/magnetars have a concurrent measurement of the ($\mathcal{Q},\tau$) parameters. For a few glitches, the ATNF catalogue provides multiple estimates of ($\mathcal{Q},\tau$). In such cases we follow once again \citet{2023MNRAS.519.5161M}, and choose the ($\mathcal{Q},\tau$) couple with the largest $\mathcal{Q}$. 
For the glitches for which we have no measurement of the healing parameter we assume $\mathcal{Q} \! = \! 1$.

Figure \ref{fig:glitch_distribution} shows the observed $\delta \nu / \nu$ as a function of the spin frequency $\nu$ of the star for observed glitches:  a seizable portion of the glitching-pulsar population is spinning in a region that makes it an interesting target for ET/CE and DECIGO/BBO.
\begin{figure}
\caption{The crosses show the spindown upper limit amplitude from Eq.~\ref{eq:h0_sd_glitch} for all glitches in the considered sample. Glitches that have a measurement of the healing parameter are indicated with thick markers. Red markers indicate magnetars; the rest are pulsars. The x-axis is gravitational wave frequency $f=2\nu$. The dashed lines show the expected smallest detectable signal with data from deci-Hz and ground-based detectors, as described in the text.}
\includegraphics[width=\columnwidth]{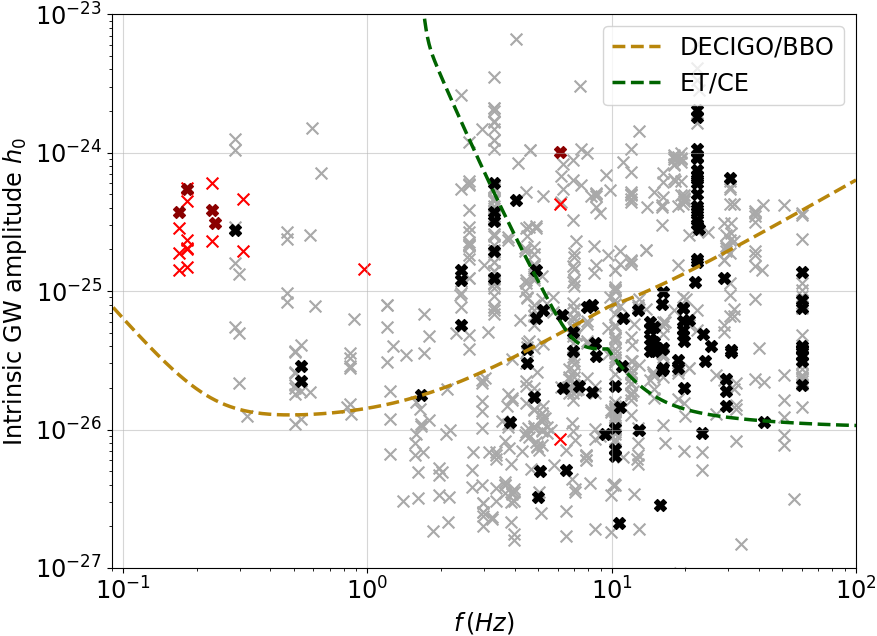}
\label{fig:glitch_results_plot}
\end{figure}

Following \cite{2023MNRAS.519.5161M} we assume that after a glitch continuous gravitational wave emission has a constant amplitude $h_0=h_{0,r}^{sd}$ and lasts only for a time $\tau$. 
Our criterion for detectability is that the signal amplitude is larger/equal to the smallest signal that we can detect (Eq.~\ref{eq:h0VsSensDepth}) : 
\begin{equation}
\label{eq:detectable_glitch}
h_{0,r}^{sd} \geq \frac{\sqrt{S_h(f)}}{\mathcal{D}} .
\end{equation}
Since $h_{0,r}^{sd} \propto 1 / \sqrt{\tau}$ (Eq.~\ref{eq:h0_sd_glitch}) and ${\mathcal{D}} \propto \sqrt{\tau}$, the relaxation time does not play any role in the detectability evaluation, but only in the estimation of the strength of the signal itself. 
For this reason we assume the same relaxation time for all glitches, and equal to 10 days. 

We adopt the sensitivity depths estimations of \cite{2023MNRAS.519.5161M} for a transient search of a 10-day continuous signal using data from the two advanced-LIGO detectors:  $\mathcal{D}^{\textrm{10d}}_{LIGO} = 28.5 \, [\textrm{Hz}]^{-1/2}$.
Accounting for the total number of effective detectors (Section \ref{sec:detectors}) we find
\begin{equation}
\begin{aligned}
\label{eq:10DaysDepth}
&\mathcal{D}^{\textrm{10d}}_{\textrm{DEC/BBO}} = 57~[\textrm{Hz}]^{-1/2}\\
&\mathcal{D}^{\textrm{10d}}_{\textrm{ET}} = 35~[\textrm{Hz}]^{-1/2}\\
&\mathcal{D}^{\textrm{10d}}_{\textrm{CE}} = 20~[\textrm{Hz}]^{-1/2} .
\end{aligned}
\end{equation}
Figure~\ref{fig:glitch_results_plot} shows that because for these searches deci-Hz detectors are competitive with the ground-based ones only for gravitational wave frequencies lower than about $6$ Hz, corresponding to spin frequencies $\nu\lesssim 3$ Hz of figure~\ref{fig:glitch_distribution}, most glitching pulsars are better detected with 3rd generation ground-based detectors ET/CE. This is in line with the results of \citet{2023MNRAS.519.5161M}.
Furthermore, the \emph{combined use} of Einstein Telescope and Cosmic Explorer would cover 61\% of all glitches with measured $\mathcal{Q}$. 

The contribution of the deci-hz detectors remains nevertheless significant. 
We find they can detect up to about 42\% of all glitches considered and about half of these are {\it{only}} detectable by deci-Hz detectors. About 38\% of glitches having a measurement of the healing parameter are detectable by deci-Hz detectors, with again about half of these exclusively from DECIGO/BBO. 
The most significant result regarding deci-Hz detectors, consists in being able to detect all but one of glitches recorded for magnetars or magnetar candidates.

\section{Discussion}
\label{sec:discussion}
We have outlined prospects for detecting continuous gravitational waves in the frequency band below 20 Hz. This frequency window will be made accessible both by 3rd generation ground-based detectors, which will be sensitive down to $\approx \, 5 \, \textrm{Hz}$, and by space-based deci-Hz detectors, which are most sensitive between 0.1 - 10 Hz.

We  focus on highly magnetised neutron stars, which are interesting for two reasons: 1) these sources can be significantly deformed due to the very strong magnetic fields they possess.
2) magnetically-induced deformations do not require ad-hoc assumptions to be realised, and the fundamental physics governing the deformation is robust.

Some indirect evidence exists that some neutron stars might be non-axisymmetric \citep{2000Natur.406..484S, 2014PhRvL.112q1102M, 2024NatAs.tmp...61D, 2024arXiv240413799M}. This comes from observing pulse periodicity in the electromagnetic spectra of these sources, consistent with them precessing, which in turn may be caused by them being non-axisymmetrically deformed with respect to the spin axis.
Of the four sources for which this phenomenon has been observed so far, three are magnetars.
This suggests that the large magnetic field magnetars might indeed generate a deformation sourcing continuous gravitational waves.

Failing to detect precession from the pulse profile of a pulsar does not mean however that the pulsar is perfectly axisymmetric. The simplest example to illustrate this is that of an orthogonal biaxial rotator, which is non-axisymmetric but would not precess. Interestingly enough, the orthogonal rotator seems to be the equilibrium configuration to which prolate deformed neutron stars tend to \citep{2002PhRvD..66h4025C}, although there is great uncertainty as to the time-scale within which this phenomenon may occur.

The high sensitivity band of current ground-based detectors falls at frequencies too high with respect to what is expected from highly magnetised neutron stars, which typically spin at frequencies below the few-Hz region. 
Indeed, recent searches for continuous gravitational waves from known pulsars \citep{2025arXiv250101495T, 2023MNRAS.519.5590C, 2022ApJ...935....1A, 2021ApJ...923...85A, 2020ApJ...902L..21A}, mostly target recycled neutron stars (in general weakly magnetised) and few mildly magnetised normal pulsars, with $B \lesssim 2 \cdot 10^{12} \, \textrm{G}$. Conversely the next generation of ground-based detectors and space-based deci-Hz detectors offer novel prospects for detection from highly magnetised sources in the low frequency range.

We show with the minimal requirement for detection, namely that the smallest detectable ellipticity is at least as small as the spin-down limit of Eq. \ref{eq:eps_sd}, all known highly magnetised pulsars or magnetars would be detectable with a fully coherent targeted search in two years of data from deci-Hz detectors.

For most of very slow ($\nu \! \leq \! 0.5 \, \textrm{Hz}$) known neutron stars for which a timing solution is available, the smallest detectable ellipticities are in all likelihood still too high to actually be plausibly realised. The band above 1 Hz in gravitational wave frequency is the most interesting, with a total of 42 known pulsars in principle detectable with an ellipticity smaller than the reference value of $\varepsilon = 1.4 \cdot 10^{-4}$, value indirectly inferred from observing precession in the magnetar 4U 0142+61. 15 of these are best detectable by deci-Hz detectors. 

Direct measurements or constraints on continuous waves emission from highly magnetised neutron stars can indirectly inform us on the internal structure and strength of magnetic fields.
In Section \ref{sec:permanent_results} we show an example of how ellipticity upper limits can be used as constraints on $\Lambda$, a quantity directly dependent on the internal magnetic field (Eq. \ref{eq:lambda_definition}).
Although this is promising, it must be emphasised that the available theoretical models rely on uncertain assumptions, making the predictions not very robust. 
Modelling magnetic deformations of neutron stars is a very difficult problem and we look forward to more theoretical work in this direction since it is crucial to guiding searches and interpreting detections and upper limits. 

The central compact objects in some supernova remnants could be hidden magnetars, for instance PSR J1852+0040 in the supernova remnant Kesteven 79. A continuous gravitational wave signal from PSR J1852+0040 could be detected by 3rd generation ground-based detectors if this source were emitting at least 20\% of its lost spin-down energy in continuous waves, or if it possessed an ellipticity of at least $7.2\cdot10^{-6}$, value compatible with an internal magnetic field in the magnetar regime.

The central compact object in Cassiopeia A, could also be a hidden magnetar. Since electromagnetic pulsations are not detected from this object, a continuous gravitational wave search needs to encompass possible values for thr frequency and frequency derivatives. 
Our estimate of a search covering the frequency band below 30 Hz and including all necessary spin-down parameters, shows that fully coherent approaches over a year are possible with searches \emph{\`a la} Einstein@Home, allowing to achieve sensitivity depths significantly higher than current searches at higher frequencies.
For instance, with data from 3rd generation detectors, at 20 Hz, the weakest detectable signal emitted by Cassiopeia A, is more than 700 times weaker than current most constraining amplitude upper limits set for this particular source at the same frequency. 
More generally, above $\approx 7$ Hz, 3rd generation detectors detectors will be able to probe ellipticities below the reference value $\approx 10^{-4}$ (see Figure \ref{fig:hidden_magnetar}). 
 
Continuous waves might play a role in the glitches of both magnetars and normal pulsars, with potentially different phenomenologies \citep{2024APh...15702921H}.
We extend the study of \citet{2023MNRAS.519.5161M} to deci-Hz frequencies and demonstrate that deci-Hz detectors are uniquely placed to detect emission from one of these categories, magnetars, covering virtually all such objects in the sample.

\section*{Acknowledgements}

G.P. thanks Masaki Ando, Seiji Kawamura, Tomohiro Ishikawa and Naoki Seto for discussions about DECIGO and its sensitivity.
G.P. thanks Brian McGloughlin and Benjamin Steltner for the fruitful discussion on directed searches' frequency coverage.
M.M. acknowledge help and fruitful discussion with Olaf Hartwig and Mauro Pieroni to extend the computation of the LISA sensitivity to the case of DECIGO.
\section*{Data Availability}

Pulsars and magnetars data is taken respectively from the ATNF \citep{Manchester_2005} and from the McGill \citep{2014ApJS..212....6O} online catalogues.
Glitch data is taken both from the ATNF \citep{Manchester_2005} and the Jodrell bank \citep{2011MNRAS.414.1679E} catalogues.



\bibliographystyle{mnras}
\bibliography{example} 




\appendix
\onecolumn
\section{Optimal signal-to-noise ratio for DECIGO/BBO}
\label{sec:appendix}
We compute the optimal signal-to-noise ratio of a matched filter output for a monochromatic signal in stationary, Gaussian, zero-mean noise, for a detector moving in a heliocentric Earth-trailing orbit.

Let us begin by defining the reference frames necessary for our calculation. We follow \citet{1997MNRAS.289..185G} both in procedure and notation, and introduce a frame $(x', y', z')$ which is stationary with respect to fixed stars.
It is convenient to define two further reference frames: the wave frame $(X, Y, Z)$, with the $Z\textrm{-axis}$ oppositely oriented to the propagation direction of the wave, and the detector's frame $(x, y, z)$, which can be conveniently chosen so that the $z\textrm{-axis}$ is orthogonal to the plane identified by the two arms of the interferometer.
The relative orientation between the latter two frames and the fixed reference frame is completely determined by six Euler angles: $(\theta, \phi, \psi)$ for the wave frame,  $(\zeta, \eta, \xi)$ for the detector frame.

The starting point of our calculation is Equation (84) of \citet{Jaranowski_1998}, which gives the optimal matched-filter signal to noise ration $d^2$:
\begin{equation}
\label{eq:jks_starting_point}
d^2 \cong \left[ \frac{1}{4} (1 + \cos^2 \iota)^2 \int_{-T_0/2}^{T_0/2} F_+^2 \,dt + \cos^2 \iota  \int_{-T_0/2}^{T_0/2} F_\times^2 \,dt\right] \frac{h_0^2}{S_h(f)} .
\end{equation}
Here we assume the wobble angle to be $\pi/2$. The \emph{inclination angle} $\iota$ is the angle between the line of sight and the total angular momentum of the star; $T_0$ is the observation time.

The space-based version of Equation \ref{eq:jks_starting_point} differs from the ground-based one because of the different time-dependency of the antenna pattern functions $F_+$ and $F_{\times}$, which is determined by the motion of the detector in space.
\citet{1997MNRAS.289..185G} calculates the antenna pattern functions for a detector orbiting the sun in a conical motion; these are
\begin{subequations}
	\begin{equation}
	\label{eq:F_plus}
	F_+ = \sin(\Gamma) \left[ A(t) \cos( 2 \xi(t) ) \cos( 2 \psi ) + B(t) \cos( 2 \xi(t) ) \sin( 2 \psi ) + C(t) \sin( 2 \xi(t) ) \cos( 2 \psi ) + D(t) \sin( 2 \xi(t) ) \sin( 2 \psi ) \right]
	\end{equation}
	\begin{equation}
	\label{eq:F_cross}
	F_{\times} = \sin(\Gamma) \left[ B(t) \cos( 2 \xi(t) ) \cos( 2 \psi ) - A(t) \cos( 2 \xi(t) ) \sin( 2 \psi ) + D(t) \sin( 2 \xi(t) ) \cos( 2 \psi ) - C(t) \sin( 2 \xi(t) ) \sin( 2 \psi ) \right]
	\end{equation}
\end{subequations}
corroborated by the time-dependent angles
\begin{equation}
\label{eq:coord_transf}
	\begin{cases}
	  \eta(t) = \eta_0 + \frac{2 \pi}{T_{orb}} t \\
	  \xi(t) = \xi_0 - \frac{2 \pi}{T_{orb}} t
	\end{cases}
\end{equation}
where $T_{orb} = 1 \, \textrm{yr}$, and $\Gamma$ is the angle between two adjacent arms of the detector. 
The functions $A(t), B(t), C(t)$ and $D(t)$ are rather complicated functions of $\zeta, \theta, \phi$ and $\eta$. For the sake of conciseness we do not show them here, but they can be found in Appendix A of \citet{1997MNRAS.289..185G}.

We can now evaluate the integrals in Equation \ref{eq:jks_starting_point} with the expression given in Eq.~\ref{eq:F_plus} and \ref{eq:F_cross}. 
Since we are interested in how space-based detectors respond to impinging gravitational waves \emph{on average}, i.e. not specifically to any sky direction or polarization, we evaluate the average optimal signal-to-noise ratio with respect to sky direction, polarization and inclination angle, as done in Equation (93) of \citet{Jaranowski_1998} \footnote{We warn the reader that our notation differs from that of \citet{Jaranowski_1998}.}.
Also, because of the 1 yr periodicity of the antenna pattern functions, for simplicity we assume $T_0$ to be either exactly a multiple of 1 year or much larger.
We calculate Equation \ref{eq:jks_starting_point} for $T_0 = 1$ yr, and average over sky position $(\phi, \theta)$, polarization ($\psi$) and inclination angle ($\iota$), obtaining
\begin{equation}
\label{eq:jks_final}
\langle d_2^2 \rangle_{\phi, \theta, \psi, \iota} = \frac{4}{25} \sin^2 \Gamma \frac{h_0^2}{S_h(f)} .
\end{equation}

The result provided in Equation \ref{eq:jks_final} is identical to that for the ground-based case, except for the geometric factor $\sin^2 \Gamma$, which in ground-based detectors, having their arms forming a right angle, equals to 1, while for DECIGO or Big Bang Observer, having two adjacent arms forming a $60^{\circ}$ angle, is $3/4$. This geometric penalty is factored-in the sensitivity estimates that we have presented by having multiplied the amplitude spectral density of space-based detectors by a factor $\sqrt{4/3}$.

\bsp	
\label{lastpage}
\end{document}